\documentclass[pre,twocolumn,superscriptaddress,showpacs]{revtex4}

\usepackage{epsfig}
\usepackage{graphicx}
\usepackage{tabularx}
\usepackage{dcolumn}  
\usepackage{bm} 
\usepackage{amsmath} 
\usepackage{times}
\usepackage{color}

\begin{document}

\title{Site and bond percolation thresholds in $K_{n,n}$-based lattices:\\
Vulnerability of quantum annealers to random qubit and coupler failures
on chimera topologies}

\author{O. Melchert}
\affiliation{Department of Physics and Astronomy, Texas A\&M University,
College Station, Texas 77843-4242, USA}

\author{Helmut G.~Katzgraber}
\affiliation{Department of Physics and Astronomy, Texas A\&M University,
College Station, Texas 77843-4242, USA}
\affiliation{Santa Fe Institute, 1399 Hyde Park Road, Santa Fe, New
Mexico 87501 USA}
\affiliation{Applied Mathematics Research Centre, Coventry University,
Coventry, CV1 5FB, England}

\author{M.~A.~Novotny}
\affiliation{Department of Physics and Astronomy, Mississippi State
University, Mississippi State, Mississippi 39762-5167, USA}
\affiliation{HPC$^2$ Center for Computational Sciences, Mississippi
State University, Mississippi State, Mississippi 39762-5167, USA}

\date{\today}

\begin{abstract}

We estimate the critical thresholds of bond and site percolation on
nonplanar, effectively two-dimensional graphs with chimera like
topology. The building blocks of these graphs are complete and symmetric
bipartite subgraphs of size $2n$, referred to as $K_{n,n}$ graphs. For
the numerical simulations we use an efficient union-find based algorithm
and employ a finite-size scaling analysis to obtain the critical
properties for both bond and site percolation.  We report the respective
percolation thresholds for different sizes of the bipartite subgraph and
verify that the associated universality class is that of standard
two-dimensional percolation. For the canonical chimera graph used in the
D-Wave Systems Inc.~quantum annealer ($n = 4$), we discuss device
failure in terms of network vulnerability, i.e., we determine the
critical fraction of qubits and couplers that can be absent due to
random failures prior to losing large-scale connectivity throughout the
device.

\end{abstract} 

\maketitle

\section{Introduction}
\label{sec:introduction}

In its most basic variant, the standard percolation model comprises a
very minimalistic model of porous media
\cite{broadbent:57,hammersley:57,hammersley:57a}. However, despite its
simplicity, percolation can be applied to problems across disciplines
ranging from forest fires to current flow in resistor networks, liquid
gelation, network connectivity, coffee brewing, simple configurational
statistics \cite{fisher:61a}, transport phenomena in ionic glasses
\cite{bunde:91}, string-bearing models that also involve a large degree
of optimization, describing, for example, vortices in high $T_c$
superconductivity \cite{pfeiffer:02,pfeiffer:03}, to name a few.
Although conceptually simple, the configurational statistics of the
percolation problem feature a nontrivial phase transition
\cite{stauffer:79,stauffer:94}. To facilitate intuition, consider, for
example, random-bond percolation on a two-dimensional square lattice
where one studies a diluted system in which only a random fraction $p$
of the edges subsist. The connected components \cite{essam:70} of the
lattice can be seen as clusters that are then analyzed with respect to
their geometric properties.  Depending on the fraction $p$ of subsisting
edges, the geometric properties of the clusters change: Exceeding a
lattice-structure dependent critical threshold $p_{\rm c}$, the model
transitions from a disconnected phase with typically small clusters to a
phase where there is a single large cluster that interconnects a finite,
nonzero fraction of the lattice sites, thus achieving large-scale
connectivity.  The appearance of this system-spanning cluster can be
described by a second-order phase transition \cite{yeomans:92}.

\begin{figure}[t!]
\begin{center}
\includegraphics[width=0.85\linewidth]{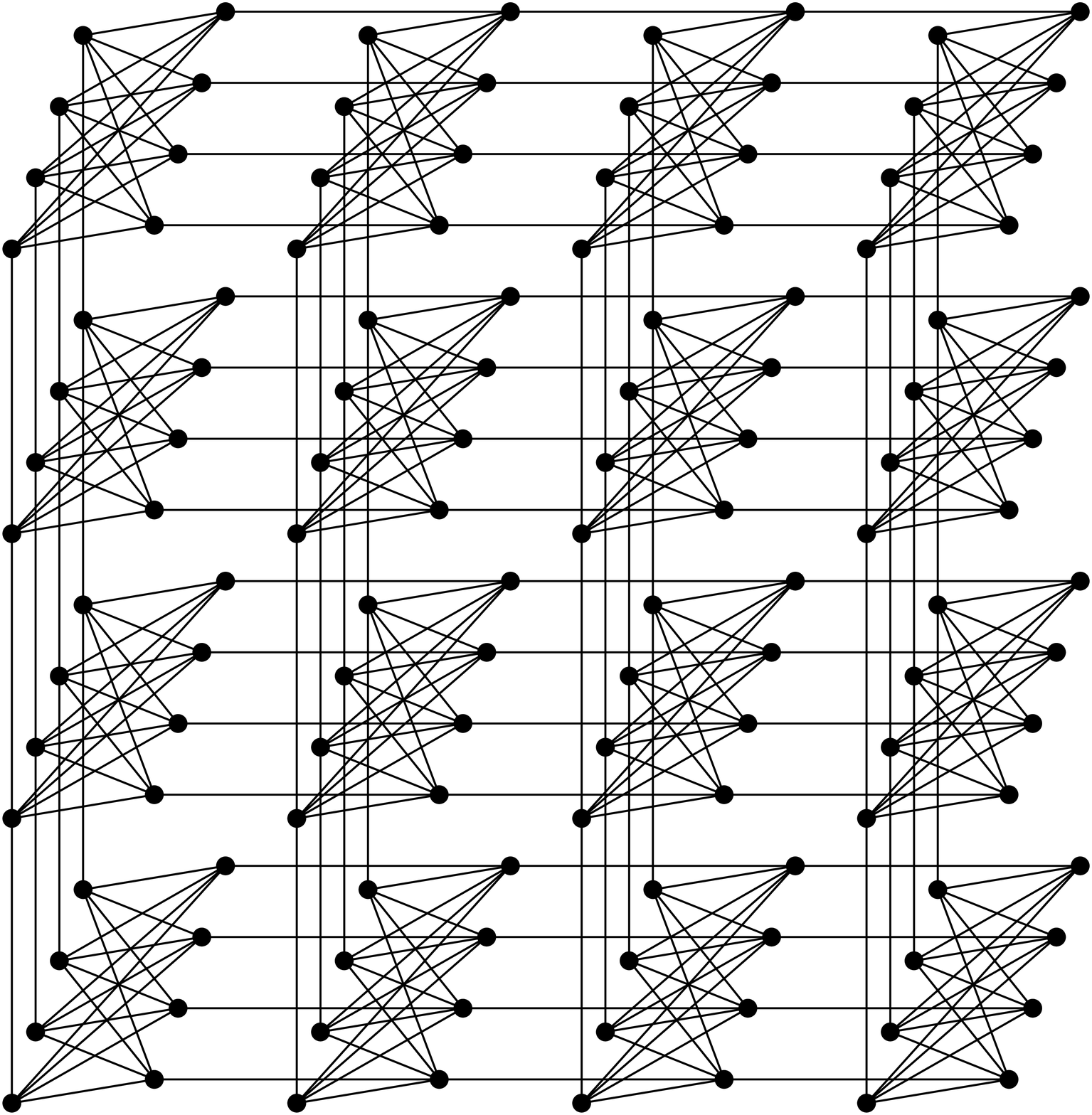}
\end{center}
\caption{
Topological representation of a chimera graph with $N=128$ sites, based
on a $4\times 4$ grid of $K_{4,4}$ subgraphs, which corresponds to the
D-Wave One Rainier quantum annealer introduced in 2011.
\label{fig:chimeraGraph}}
\end{figure}

Because the location of the percolation critical point is sensitive to
the topology of the underlying graph, there is general interest in
understanding these threshold values for relevant model systems
\cite{becker:09,saberi:15}. In some cases it is possible to derive these
thresholds exactly by analytical calculations. For example, in
Refs.~\cite{newman:01a} and \cite{callaway:00} a generating function
approach was developed to determine the statistical properties of random
graphs with arbitrary degree distribution (e.g., Erd\H{o}s-R\'{e}nyi
random graph ensembles). Unfortunately, this is only typically possible
for few exceptional cases and so it is generally necessary to rely on
numerical approaches (e.g., via Monte Carlo simulations) to calculate
the precise percolation thresholds via a finite-size scaling analysis on
finite lattices.  In this regard, from a point of view of numerical
simulations, significant algorithmic progress has been made by using
bookkeeping concepts based on union-find data structures
\cite{cormen:01} that led to highly efficient algorithms for bond and
site percolation problems \cite{ziff:00,ziff:01}. For an extension of
the algorithmic procedure to continuum percolation models, describing
spatially extended, randomly oriented and possibly overlapping objects,
see Ref.~\cite{mertens:12}.

Here, we perform numerical simulations to estimate the thresholds for
both bond and site percolation on nonplanar effectively two-dimensional
lattices, where the elementary building blocks are given by $K_{n,n}$
subgraphs, i.e., complete bipartite subgraphs of size $2\times n$
\cite{essam:70} (see Sec.~\ref{sec:model} below for details).  The
particular choice of $n=4$ is known as the chimera graph
\cite{bunyk:14}, which is the native (hardware) topology of the
special-purpose quantum annealing device developed by D-Wave Systems
Inc.~\cite{comment:d-wave}.  Our motivation to study percolation on the
chimera graph stems from the possible existence of fabrication defects
or trapped fluxes that might lead to either malfunctioning qubits (see,
for example, Fig.~1 in Ref.~\cite{bian:14}) or couplers, thus
restricting the size of embeddable problems on the D-Wave chip
\cite{klymko:14}.  From an alternative point of view, adopted in the
context of network robustness and vulnerability
\cite{albert:00,callaway:00}, the fraction $f < f_{\rm c}=1-p_{\rm c}$
might be interpreted as the fraction of sites or bonds that might be
absent due to random failures, such as fabrication defects, trapped
fluxes, or operational errors, while still maintaining large-scale
connectivity throughout the chip. Above $f_{\rm c}$, however,
large-scale connectivity will be lost, leaving small-sized
interconnected qubit clusters only. This could also affect the
functionality of the chip and become an important issue for particular
embeddings of problems where a large fraction of (randomly chosen)
couplers are turned off \cite{comment:vinci}.

There are multiple reasons to compute the percolation threshold of
chimera like lattices: First, the {\em native} \cite{comment:native}
benchmark problem to study the D-Wave device is an Ising spin glass
\cite{binder:86,stein:13} on the chimera lattice. Because true optima
need to be computed using classical simulation techniques to verify that
the device can, indeed, find the solutions of the problems, efficient
optimization techniques have to be used
\cite{katzgraber:14,katzgraber:15}. Often, not only is the minimum of
the cost function needed, but also the ground-state degeneracy.
Monte Carlo based methods, such as isoenergetic cluster moves
\cite{zhu:15b}, have proven to be extremely efficient in studying systems
with low ground-state degeneracy; however, to improve the efficiency of
the algorithm, it is imperative to know the site percolation threshold
of the underlying lattice.  Simple subgraphs with known ground states,
such as one-dimensional graphs \cite{hen:15a,king:15} and spanning trees
\cite{hall:15}, have been investigated on the D-Wave device. In
addition, there have been attempts to create hard benchmark problems
using planted solutions \cite{boixo:14}.  While these elegant approaches
have the advantage that the solution to the problem to be optimized is
known {\em a priori}, the used construction procedures might lead to diluted
graphs in which only a finite fraction of edges on the lattice are used.
Although the construction procedure contains correlations and the adding
of edges is not purely random, the problem shares characteristics of
random bond percolation and so disconnected clusters might occur.
Finally, next-generation hardware might likely include a more
interconnected topology, i.e., larger values of $n$ in the $K_{n,n}$
building blocks.  Understanding the possible failure rate of these more
complex architectures due to percolation is of great importance in the
design and scalability of future-generation devices.

Here, we numerically study the $K_{4,4}$-based chimera lattice with up
to $N \approx 20\,000$ sites and estimate the site-percolation threshold
by performing a finite-size scaling of the Binder parameter
\cite{binder:81} to be $p_{\rm c} \approx 0.3866(3)$ (see also the
Supplemental Material of Ref.~\cite{zhu:15b}).  In addition, we study
general $K_{n,n}$-based chimera like lattices with $n = 2, \ldots, 8$
and estimate the corresponding bond- and site-percolation thresholds
$p_{{\rm c},n}$, as well as the associated critical exponents that
describe the percolation transition.

The paper is organized as follows.  In Sec.~\ref{sec:model} we introduce
chimera graphs in more detail, followed by details of the simulations in
Sec.~\ref{sec:num} and results in Sec.~\ref{sec:results}.  We summarize
and discuss our findings in Sec.~\ref{sec:summary}.

\section{The chimera topology\label{sec:model}}

We consider nonplanar, effectively two-dimensional lattice graphs
$G=(V,E)$, consisting of a vertex set $V$, containing $N\equiv v(G)$
vertices, and an edge set $E$, containing $M\equiv e(G)$ undirected
edges. The elementary building blocks of these graphs are $K_{n,n}$
subgraphs, i.e., complete bipartite graphs \cite{essam:70,comment:essam}
containing $2\times n$ sites. These subgraphs can be partitioned into
two vertex subsets $V_1$ and $V_2$ of size $v_1=v_2=n$ and have an edge
set, consisting of all possible $v_1\times v_2$ undirected edges with
one terminal vertex in $V_1$ and one in $V_2$.

To compose the full chimera graph $G$ with $N=2\times n \times L_x\times
L_y$ vertices, $K_{n,n}$ subgraphs are arranged on a $L_x \times L_y$
grid.  For horizontally (vertically) adjacent subgraphs $K_{n,n}$ and
$K^{\prime}_{n,n}$, and following an ordering of the vertices in the
respective vertex subsets $V_{1,2}$ and $V^{\prime}_{1,2}$, vertices out
of $V_1$ ($V_2$) are joined to their respective mirror vertex in
$V^{\prime}_1$ ($V^{\prime}_2$).  The particular choice with $n=4$
yields the canonical chimera graph. A topological representation of such
a chimera graph with $L_x=L_y=4$ is shown in
Fig.~\ref{fig:chimeraGraph}.

Subsequently, we consider chimera like graphs of size $N=8192$ ($L_x =
L_y = 32$) up to $N=294912$ ($L_x = L_y = 192$) in order to perform a
finite-size scaling analysis for different subgraph sizes and to
determine the respective thresholds for bond, as well as site
percolation.  Note that there is a difference between the practical
(small) graph sizes to which the D-Wave chip architecture is currently
limited to (see Ref.~\cite{comment:d-wave}), as opposed to large systems
that, from a point of view of statistical physics, display a decent
finite-size scaling behavior. Given that between 2011 and 2015 the
number of sites increased from $N = 128$ (Rainier chip, see
Fig.~\ref{fig:chimeraGraph}) to $N = 1152$ (Washington chip) on the
D-Wave device, we can expect \cite{comment:dream} to see chips of the
order of sites studied in this work by 2019.

\begin{figure}[t!]
\begin{center}
\includegraphics[width=1.0\linewidth]{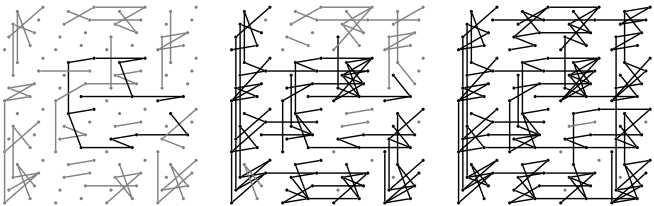}
\end{center}
\caption{
Instances of bond percolation configurations on a chimera graph with
$N=128$ sites. From left to right: $p=0.2$, $p=0.38\approx p_{\rm c}$,
and, $p=0.42$. The vertices and edges belonging to the largest connected
component are colored black and the remaining vertices and subsisting
edges are colored gray.
\label{fig:chimeraGraph_bondPerc}}
\end{figure}

\section{Numerical details\label{sec:num}}

For the numerical simulations we use the highly efficient algorithm by
Newman and Ziff \cite{ziff:00,ziff:01} based on a union-find data
structure \cite{cormen:01}. In particular, we implemented union by rank
and path compression for the find-part of the bookkeeping procedure.

Within the bond-percolation study, one sweep of the algorithm goes as
follows: First, a random permutation of the edges in the edge set $E$ of
$G$ is obtained by means of a Fisher-Yates shuffle \cite{cormen:01}
[having algorithmic complexity ${\mathcal O}(M)$ with $M$ the number of
edges].  Initially, each vertex is its own single-site cluster. Edges
from the shuffled edge set are added one at a time and for each edge it
is checked whether its incident vertices belong to different clusters.
If this is the case, the respective clusters are merged using the
union-by-rank approach. Once all edges have been probed, one lattice
sweep is completed.  We measure the size of the largest cluster and the
average size of all finite clusters. Because of the previously described
approach, these can be measured very efficiently with a resolution of
$\Delta p = 1/M$.  However, to keep the amount of raw data manageable, we
consider only approximately $80$ values of $p$ in the vicinity of the
critical point. Error bars are computed by averaging over $5\times 10^4$
sweeps for each system size studied.

Note that while the bond percolation variant of the algorithm only
requires an edge list representing $E$---i.e., the edge set of the
underlying graph---the site percolation variant of the algorithm relies
on an adjacency list of $G$, i.e., a collection of lists of neighbors
for each node \cite{cormen:01}.

\begin{table}[!tb]
\caption{\label{tab:tab1}
Critical parameters of bond percolation (BP) and site percolation (SP)
for $n = 4$ chimera graphs. From left to right: Critical percolation
threshold $p_{\rm c}$, critical exponents $\nu$ and $\beta$ (obtained
from a finite-size scaling of the order parameter), as well as $\gamma$
(obtained from the order parameter fluctuations and the scaling behavior
of the average size of the finite clusters). For details see the main
text.}
\begin{tabular*}{\columnwidth}{@{\extracolsep{\fill}} l l l l l l l}
\hline
\hline
Type & $p_{\rm c}$  & $\nu$ & $\beta$ & $\gamma$ \\
\hline
BP  & 0.2943(1)   & 1.34(2) & 0.146(8) & 2.42(2) \\
SP  & 0.38722(7)  & 1.34(3) & 0.145(5) & 2.41(2) \\
\hline
\hline
\end{tabular*}
\end{table}

\section{Results \label{sec:results}}

We illustrate our approach and data analysis in detail using a
finite-size scaling analysis of the canonical $K_{4,4}$-based chimera
lattice. However, we have performed the same algorithm for all $K_{n,n}$
lattices with $n = 2, \ldots, 8$.

\subsection{Bond percolation on $n = 4$ chimera graphs}
\label{subsec:res_bondPerc}

The observables we consider can be rescaled following a generic scaling
assumption, i.e.,
\begin{eqnarray}
y(p,N)= N^{-b/2}~f[(p-p_{\rm c}) N^{1/(2\nu)}], \label{eq:scalingAssumption}
\end{eqnarray}
where $\nu$ and $b$ represent dimensionless critical exponents (or
ratios thereof, see below), $p_{\rm c}$ is the critical threshold, and
$f[\cdot]$ denotes an unknown scaling function
\cite{stauffer:94,binder:02a}. Following
Eq.~\eqref{eq:scalingAssumption}, data curves of the observable $y(p,N)$
computed at different values of $p$ and $N$ fall on top of each other,
if the scaling parameters $p_{\rm c}$, $\nu$, and $b$ are chosen
properly. The values of the scaling parameters that yield the best data
collapse determine the numerical values of the critical point and the
critical exponents that govern the behavior of the underlying observable
$y(p,N)$.

To determine the optimal data collapse for a given set of data curves we
perform a computer-assisted scaling analysis
\cite{melchert-autoscale:09,comment:sorge}.  Here, the ``quality'' of
the data collapse is measured by the mean-square distance of the data
points to the master scaling curve $S$, described by the scaling
function, in units of the standard error of the data points
\cite{houdayer:04}. It is a quantitative measure for the quality of a
data collapse that is far superior than the commonly used eyeballing
scaling analysis.  It is common practice to limit the analysis to the
larger system sizes, for which corrections to scaling are less
pronounced, and to discard small system sizes that are typically
affected by stronger systematic corrections to scaling
\cite{binder:02a}.  In general, systematic corrections to scaling result
in a scaling behavior that deviates from that predicted by the scaling
assumption, Eq.~\eqref{eq:scalingAssumption}.  Note that such corrections
are not taken into account here. Furthermore, while $S$ can be
influenced by potential corrections to scaling, it might not be
interpreted as a measure for these corrections.  Here, if not stated
explicitly, the scaling analysis is limited to the three largest systems
simulated.

Example instances of bond-percolation configurations in the subcritical,
critical, and supercritical regimes for chimera graphs with $N=128$ sites
are shown in Fig.~\ref{fig:chimeraGraph_bondPerc}.  The resulting
numerical estimates of the critical percolation thresholds and
corresponding critical exponents for bond and site percolation are
listed in Table \ref{tab:tab1}.

\subsubsection{Analysis of the Binder ratio}\label{par:BinderPar}

First we consider the relative size of the largest cluster of connected
vertices $s_{\rm max}$.  The dimensionless ratio, known as the Binder
parameter \cite{binder:81b}, is defined via
\begin{eqnarray}
b(p) = 
    \frac{1}{2} 
    \left[3 
	- \frac{\langle s_{\rm max}^4(p) \rangle}{ \langle s_{\rm max}^2(p) \rangle^2} 
    \right] .   
\label{eq:binderPar}
\end{eqnarray}
Here, $\langle \cdots \rangle$ represents an average over sweeps.
Because the system-size-dependent part of the scaling function in
Eq.~\eqref{eq:scalingAssumption} cancels out in the Binder ratio, it has
a simple scaling form that follows Eq.~\eqref{eq:scalingAssumption} with
$b = 0$. When $p = p_{\rm c}$ the argument of the scaling function $f$
is zero and thus system-size independent. This means that data for
different system sizes $N$ cross at $p = p_{\rm c}$ [see inset to
Fig.~\ref{fig:orderPar_fss}(a)].  Determining the correct thermodynamic
values of $p_{\rm c}$ and $\nu$ results in a data collapse, as can be
seen in the main panel of Fig.~\ref{fig:orderPar_fss}(a).  There are
visible corrections to scaling in the nonpercolating phase, i.e., for
$p<p_{\rm c}$.  To account for this, the scaling analysis is performed
in the interval $\epsilon \in [-0.25,1.75]$ on the rescaled $p$ axis to
accentuate the region where $b(p)$ scales well.  Consequently, the best
data collapse yields $p_{\rm c}=0.2946(2)$ and $\nu=1.34(2)$ with a
quality $S=1.10$ of the data collapse \cite{comment:S}. Note that the
numerical value of the correlation length exponent $\nu$ is in good
agreement with $\nu=4/3\approx 1.333$, the standard value for
percolation in two-dimensional lattices.

\subsubsection{Analysis of the order parameter}\label{par:orderPar}

The scaling of the disorder-averaged order parameter
\begin{eqnarray}
P_{\rm max}(p)=\langle s_{\rm max}(p) \rangle \label{eq:orderPar}
\end{eqnarray}
is expected to follow Eq.~\eqref{eq:scalingAssumption} with
$b=\beta/\nu$.  Here, $\beta$ refers to the percolation strength
exponent that governs the growth of the largest cluster with increasing
system size at fixed $p=p_{\rm c}$. The best data collapse (obtained in
the range $\epsilon \in [-0.5,0.5]$) yields $p_{\rm c}=0.2943(1)$,
$\nu=1.37(4)$, and $\beta=0.146(8)$ with a quality $S=1.10$ [see
Fig.~\ref{fig:orderPar_fss}(b)]. If we fix the numerical values of the
critical exponents to their exact values for two-dimensional percolation
($\nu=4/3\approx 1.333$ and $\beta=5/36\approx0.139$) we are left with
only one adjustable parameter, resulting in the estimate $p_{\rm
c}=0.2944(6)$ with (expectedly worse) collapse quality $S=4.5$. However,
both numerical values are still in good agreement.

\subsubsection{Analysis of the order parameter
fluctuations}\label{par:orderParFluct}

\begin{figure}[th!p]
\begin{center}
\includegraphics[width=1.0\linewidth]{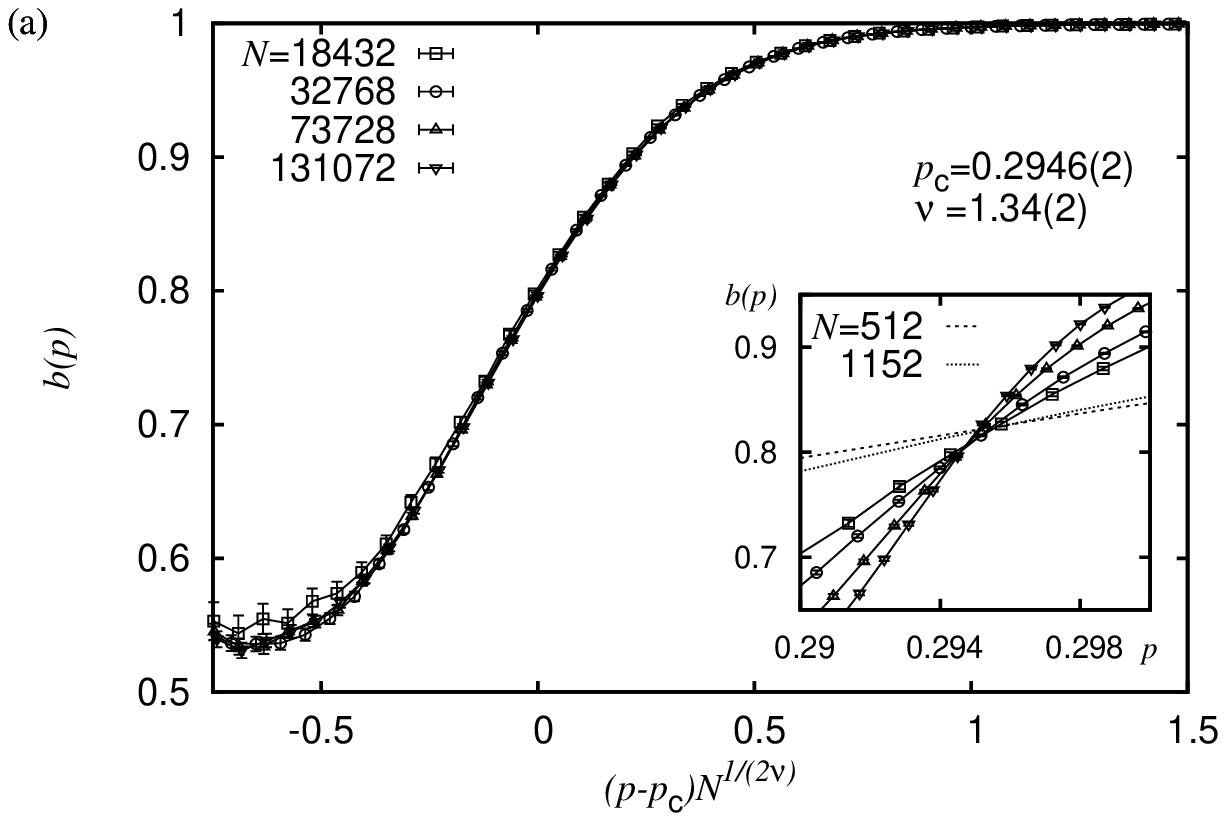}
\includegraphics[width=1.0\linewidth]{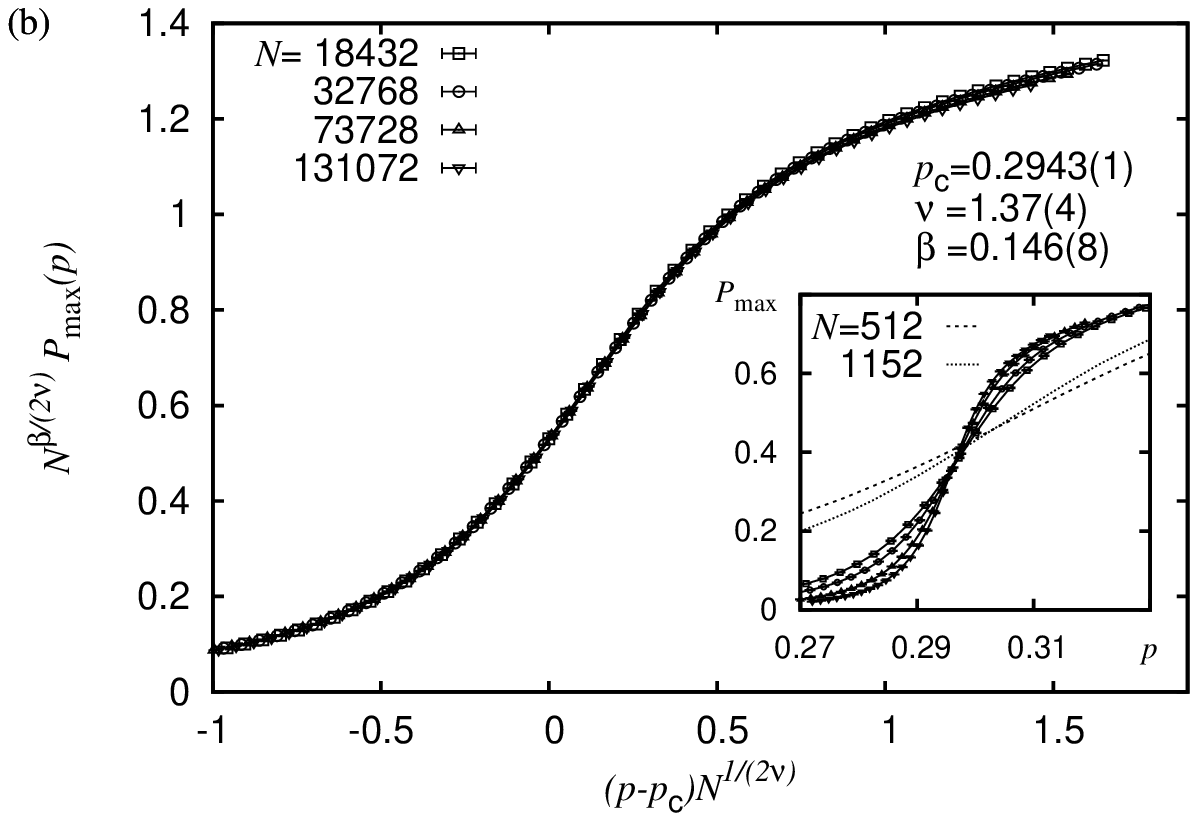}
\includegraphics[width=1.0\linewidth]{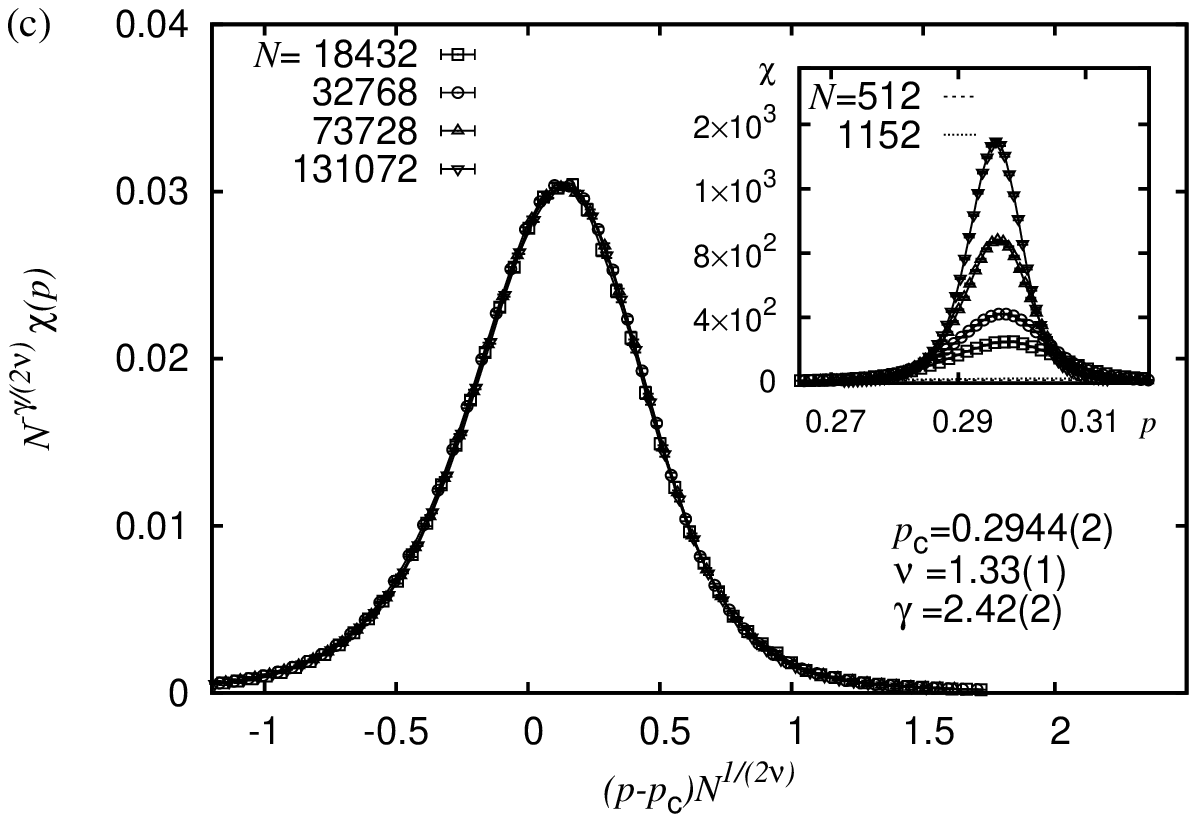}
\end{center}
\caption{
Finite-size scaling analysis of the relative size $s_{\rm max}$ of the
largest cluster of sites for the bond-percolation problem on chimera
graphs. The main panels always show the scaled data according to
Eq.~\eqref{eq:scalingAssumption}, whereas the insets display the
unscaled data in the vicinity of the critical point.  (a) Binder ratio
$b(p)$, (b) disorder-averaged order parameter $P_{\rm max}(p)$, and (c)
fluctuation $\chi(p)=N\times{\rm var}(s_{\rm max})$ of the order
parameter.  
Note that the insets feature two additional data curves, illustrating the
statistical properties of small chimera graphs of current quantum annealing
machines with $N = 512$ and $1152$ qubits (sites).
current quantum annealing machines with $N=512$ $N=1152$ qubits/sites.
\label{fig:orderPar_fss}}
\end{figure}

A third critical exponent can be estimated from the scaling of the order
parameter fluctuations $\chi(p)$, i.e.,
\begin{eqnarray}
\chi(p)= N [ \langle s_{\rm max}^2(p) \rangle - \langle s_{\rm max}(p) \rangle^2  ]. \label{eq:suscept}
\end{eqnarray}
The fluctuations $\chi(p)$ are expected to scale according to
Eq.~\eqref{eq:scalingAssumption} allowing one to determine the
fluctuation exponent $\gamma$ through $b=-\gamma/\nu$. Here, so as to
perform the best possible data collapse, the nonsymmetric range
$\epsilon \in [-0.3,1.0]$ is chosen. This is motivated by the
observation that the peaks of the data curves are located in the
superpercolating regime, with the precise location of the peaks
approaching their asymptotic value from above. Hence the aforementioned
asymmetric interval accentuates the region around the peaks, resulting
in the estimates $p_{\rm c}=0.2944(2)$, $\nu=1.33(1)$, and
$\gamma=2.42(2)$ with a quality $S=0.74$ [see
Fig.~\ref{fig:orderPar_fss}(c) for a scaling collapse].  Note that the
numerical value of the fluctuation exponent is in reasonable agreement
with the expected exact value for two-dimensional percolation, namely,
$\gamma=43/18\approx 2.389$.

\subsection{Site percolation on chimera graphs}
\label{subsec:res_sitePerc}

The analysis of the site-percolation problem is analogous to the
analysis performed for bond percolation
(Sec.~\ref{subsec:res_bondPerc}).  Note that, as discussed in
Ref.~\cite{zhu:15b}, the location of the site-percolation threshold is
pivotal for the efficient and correct performance of cluster algorithms
designed to simulate spin-glass models in arbitrary space dimensions.
In Ref.~\cite{zhu:15b}, the authors simulated chimera lattices with up
to $N=20\,000$ sites, and estimated the site-percolation threshold from
the finite-size scaling of the Binder parameter, finding $p_{\rm
c}\approx 0.3866(3)$ with $\nu=1.39(1)$.

We perform an analysis of the order parameter using systems of up to
$N=294912 = 8\times192^2$ sites.  By increasing the system sizes by
approximately one order of magnitude in comparison to the study of
Ref.~\cite{zhu:15b} we are able to verify that the exponent $\nu$ is
very likely in the two-dimensional percolation universality class.  From
an analysis of the Binder ratio we obtain $p_{\rm c}=0.3871(1)$, which,
compared to the estimate of Ref.~\cite{zhu:15b}, turns out to be
slightly larger.

Although the associated critical exponent $\nu=1.33(2)$ is in good
agreement with the two-dimensional percolation value, the data-collapse
quality $S=3.73$ is rather large, reflecting that there are deviations
from the expected scaling behavior, similar to the difficulties
encountered in the analysis of bond percolation in
Sec.~\ref{subsec:res_bondPerc}.

To ensure that our analysis of the order parameter and its fluctuations
is as precise as possible, we increased the number of samples studied to
$5 \times 10^5$.  Our estimates of the critical parameter for site
percolation on the $K_{4,4}$-based chimera lattice are $p_{\rm
c}=0.38722(7)$, $\nu=1.34(3)$, $\beta=0.145(5)$ ($\epsilon =
[-0.20:0.20]$; $S=1.00$).  Furthermore, the parameter estimates obtained
from the order parameter fluctuations are $p_{\rm c}=0.3870(2)$,
$\nu=1.34(1)$, and $\gamma=2.41(2)$ ($\epsilon = [-0.70:0.70]$;
$S=2.50$).  Note that both estimates of $p_{{\rm c}}$ are in agreement
with each other and in agreement with the Binder cumulant values
estimated above.  In both cases, the critical exponent $\nu$ is in
agreement with the exact value of two-dimensional percolation and
$\beta$ and $\gamma$ are in reasonable agreement with their exact
two-dimensional values (i.e., within two standard deviations).  Despite
the numerical values of $\beta$ and $\gamma$ not matching the known
values of two-dimensional ($2$D) percolation exactly, we believe, based
on the other exponents and our general expectations on this short-ranged
percolation model, that the transition is actually of the universality
class of $2$D random percolation.

\begin{table}[b!]
\caption{\label{tab:tab2}
Percolation thresholds on generalized chimera graphs built from
$K_{n,n}$ elementary cells of size $n=2$ --- $8$. From left to
right: size $n$ of the $K_{n,n}$ elementary cell (each cell contains
$2n$ sites), critical points $p_{{\rm c},n}$ obtained from an analysis
of the order parameter for bond percolation (BP) and site percolation
(SP), respectively.
} 
\begin{ruledtabular}
\begin{tabular}[c]{l@{\quad}llll}
  $n$ & $p_{{\rm c},n}$ (BP)  & $p_{{\rm c},n}$ (SP)\\
\hline
2  & 0.44778(15)    & 0.51294(7)  \\
3  & 0.35502(15)    & 0.43760(15)  \\
4  & 0.29427(12)    & 0.38675(7)   \\
5  & 0.25159(13)    & 0.35115(13)  \\
6  & 0.21942(11)    & 0.32232(13)  \\
7  & 0.19475(9)     & 0.30052(14)  \\
8  & 0.17496(10)    & 0.28103(11)  \\
\end{tabular}
\end{ruledtabular}
\end{table}

\subsection{Percolation thresholds on generalized chimera graphs}
\label{subsec:res_generalized}

\begin{figure}[t!]
\begin{center}
\includegraphics[width=1.0\linewidth]{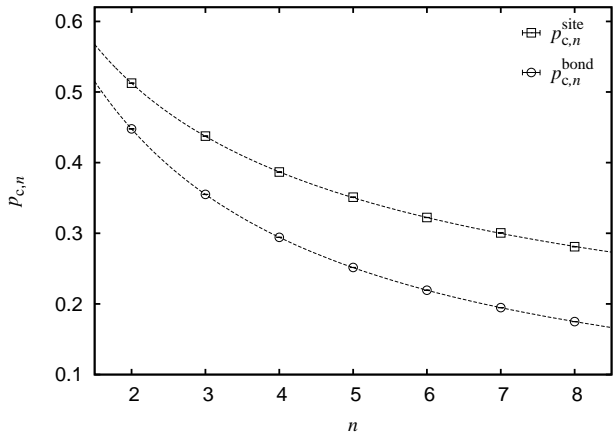}
\end{center}
\caption{
Bond and site percolation thresholds for chimera like graphs built from
$K_{n,n}$ subgraphs. The percolation thresholds decrease with increasing
average vertex degree and thus of $n$, the cell size.  The dashed lines
represent fits to functions of the form $f(n)=a(n-\Delta n)^{-b}$
(see text for details).
\label{fig:pc_genChim}}
\end{figure}

For $K_{n,n}$-based generalized chimera graphs one might intuitively
expect that the percolation threshold is a decreasing function of the
average vertex degree and thus of $n$ (however, note that
counterexamples can be constructed \cite{wierman:02} on planar
lattices).  Here, we perform a finite-size scaling analysis for the
disorder-averaged relative size of the largest cluster, i.e., the order
parameter [Eq.~(\ref{eq:orderPar})], to determine the thresholds for
$n=2$ --- $8$ (the standard chimera graph has $n=4$).  Therefore, for
each value of $n$, we consider three system sizes with up to $N=131044$
sites (the precise value of $N$ depends of the choice of $n$, of
course). Furthermore, we consider $10^4$ different permutations of the
edge set or the vertex set for both bond and site percolation to compute
$\langle s_{\rm max}(p) \rangle$.  As can be seen in
Fig.~\ref{fig:pc_genChim}, the thresholds decrease with increasing $n$
and can be fit well by functions of the form $f(n)=a(n-\Delta n)^{-b}$.
In this regard we find $a={\mathcal O}(1)$, $\Delta n= {\mathcal O}(1)$
and $b\approx 1$ for bond percolation and $b\approx 0.5$ for site
percolation.  For the bond-percolation variant one might further
rephrase this scaling in terms of the number of internal $K_{n,n}$
edges, i.e., $m=n^2$, to also find a scaling with a characteristic
exponent $b\approx 0.5$. In either case, this suggests that in the
asymptotic limit, $p_{\rm c} \to 0$ as $n \to \infty$. The results of
the finite-size scaling analysis are listed in Table \ref{tab:tab2} (the
results for the canonical chimera lattice are again listed for $n=4$).

The quality of the data collapse is somewhat sensitive to the scaling
interval $\epsilon$ chosen in the course of the analysis.  For example,
for the critical point $p_{{\rm c},2}$ for site percolation on the
$K_{2,2}$ chimera graph we obtained estimates in the range $p_{{\rm
c},2} = 0.5124(1)$ ($\epsilon = [-1.00:0.75]$; $S=1.54$) to $p_{{\rm
c},2} = 0.5129(2)$ ($\epsilon = [-0.50:0.50]$; $S=1.97$).  Generally, we
expect that a narrower scaling interval $\epsilon$---enclosing the
critical point without extending too far into the off-critical region
where deviations from the scaling behavior are expected---should lead to
a more reliable estimate of $p_{{\rm c}}$. For example, for the given
statistics (e.g., $10^4$ samples), restricting the scaling interval
further to the range $\epsilon = [-0.20:0.30]$ results in $p_{{\rm c},2}
= 0.51301(15)$, $\nu=1.32(5)$, and $\beta=0.145(7)$ ($S=1.84$).  The
scaling exponents are also in agreement with the exact two-dimensional
values. Increasing the statistics by a factor of $10$ to $10^5$
independent samples effectively allows us to add one digit of precision,
i.e., $p_{{\rm c},2} = 0.51294(7)$ ($\epsilon = [-0.30:0.30]$;
$S=0.30$), a result that is in good agreement with an independent
estimate by Ziff \cite{comment:rziff_pc2}.

What does this mean for architectures built from $K_{n,n}$ subgraphs?
From a point of view of network robustness and vulnerability, increasing
$n$ leads to a hardware topology that is less vulnerable to a random
failure of qubits.  For example, while the native D-Wave design with $n=4$
allows for a random failure of approximately $62\%$ of the qubits
($70\%$ of the couplers) without losing large-scale connectivity, this
value rises to about $72\%$ ($83\%$ in the case of couplers) if the size of
the elementary building blocks is scaled up only by a factor of $2$ to
$n=8$. Therefore, using topologies that have high connectivity or, for
example, small-world properties \cite{guclu:06} is key in designing
quantum annealing machines robust to random failures of qubits and
couplers.

\begin{figure}[th!p]
\begin{center}
\includegraphics[width=1.0\linewidth]{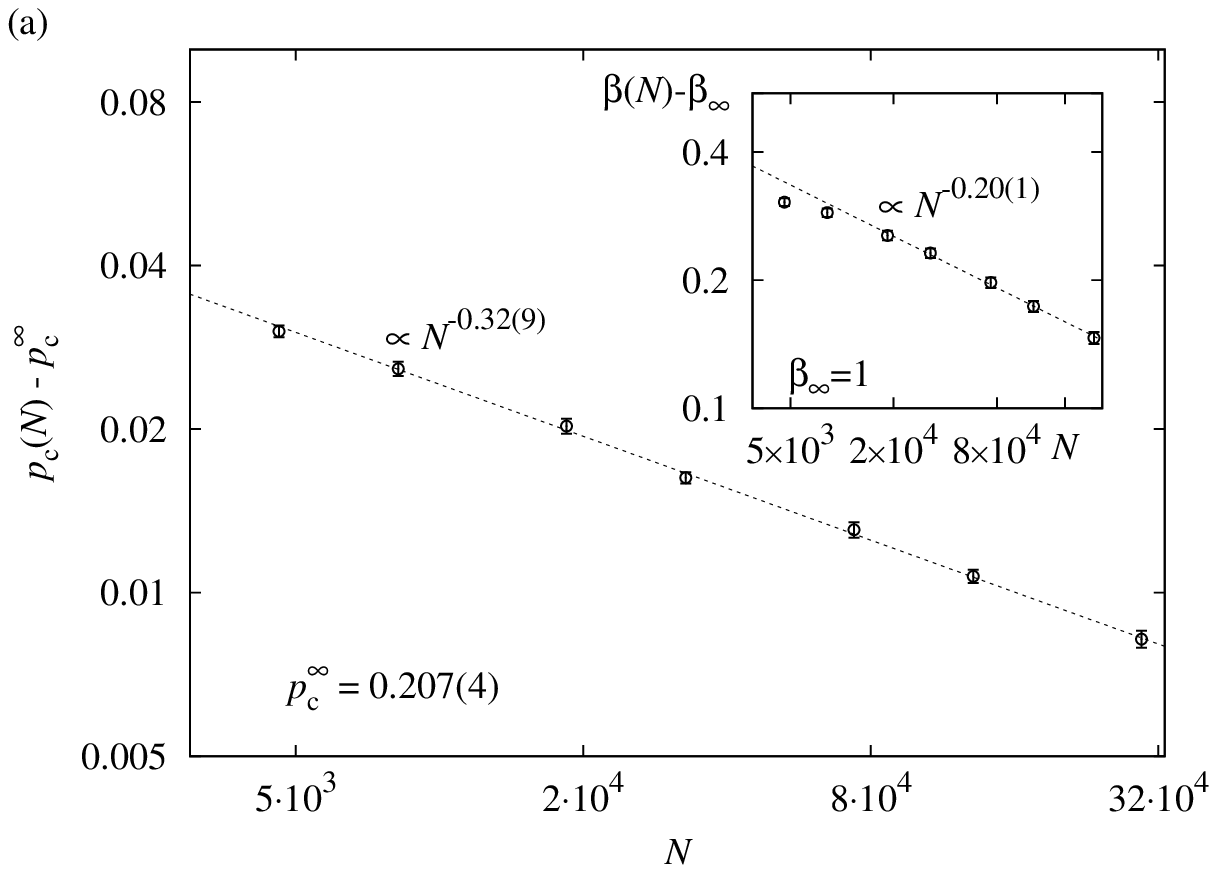}
\includegraphics[width=1.0\linewidth]{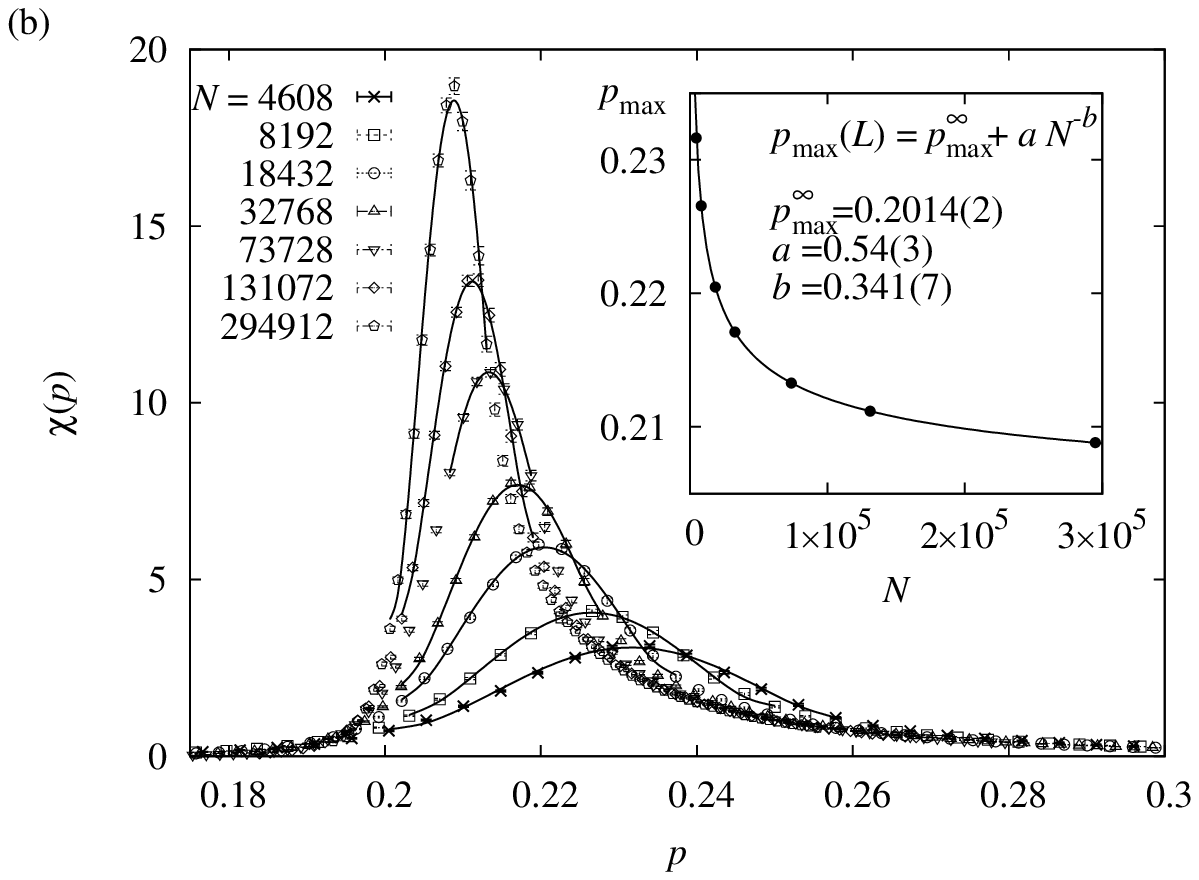}
\end{center}
\caption{
Finite-size scaling analysis of the relative size $s_{\rm max}$ of the
largest cluster of sites for the site-percolation problem on the
small-world enhanced chimera graphs (see
Sec.~\ref{sec:smallWorldChimera}). (a) Scaling of the effective
system-size-dependent estimate of $p_{\rm c}(N)$. The inset shows a
scaling of $\beta(N)$, as discussed in the text. (b) Fit of
fifth-order polynomials to the finite-size fluctuations to estimate the
system-size-dependent peak positions $p_{\rm max}(N)$. The inset shows
the extrapolation to the asymptotic critical point $p_{\rm
max}^{\infty}$, as discussed in the text.
\label{fig:SWCG_fss}}
\end{figure}  

\subsection{Small-world enhanced chimera graphs} 
\label{sec:smallWorldChimera}

We now discuss how to improve the stability of chimera like lattices by
merely increasing the average degree by one via the addition of $N/2$
``small-world'' (SW) bonds to the existing regular chimera graph.  This
results in a supergraph $G^\prime$ of $G$, which we refer to as a
small-world chimera graph (SWCG).  Our aim is to determine the location
of the site-percolation threshold for the ensemble of SWCGs and to
assess the gain in network robustness. The additional SW bonds that make
up an instance of a SWCG are obtained by the following three-step
procedure: (i) generate a list of $N$ integers that represent the
vertices of the (plain) chimera graph, (ii) obtain a random permutation
of the list, and (iii) interpret subsequent pairs of integers as the
end vertices of $N/2$ additional bonds that, in turn, are added to the
initial graph.  In doing so, the degree of each vertex increases by {\em
exactly} one \cite{comment:odd}. The resulting percolation thresholds
can be expected to decrease with decreasing average degree, and,
consequently the ensemble of SWCGs can be expected to be less
vulnerable to random qubit failures. This is in agreement with the
containment principle due to Fisher \cite{fisher:61a}, stating that if
$G$ results from $G^\prime$ by removing a fraction of its bonds (i.e.,
$G$ being a spanning subgraph of $G^\prime$; see Ref.~\cite{essam:70}),
then $p_{\rm c}^{G^\prime} \leq p_{\rm c}^G$ for {\em both} bond and
site percolation.

For the SWCGs, it is anticipated that there is a scaling window around
$p_{\rm c}$ that has mean-field exponents. A proof of such scaling
window exists on quasi-random graphs \cite{nachmias:09}.  Figure
\ref{fig:SWCG_fss} illustrates a finite-size scaling analysis of the
order parameter and its associated finite-size susceptibility for the
site-percolation problem on SWCGs. In the vicinity of the critical point
we expect the unscaled order parameter data to scale as
\begin{eqnarray}
P_{\rm max}(p) \sim |p-p_{\rm c}|^{\beta}. \label{eq:orderParAnalysis_SWCG}
\end{eqnarray}
From the data corresponding to different system sizes, we obtain the
system-size-dependent effective estimates $p_{\rm c}(N)$ and $\beta(N)$.
From the effective critical points we extrapolate to the asymptotic
critical point $p_{\rm c}^\infty$ by fitting the data to
\begin{eqnarray}
p_{\rm c} = p_{\rm c}^{\infty} + a N^{-b}, 
\end{eqnarray}
with $p_{\rm c}^{\infty} = 0.207(4)$, $a = 0.5(3)$, and $b = 0.32(9)$,
as shown in the main plot of Fig.~\ref{fig:SWCG_fss}(a).  Similarly, the
sequence of exponents $\beta(N)$ is fit well by
\begin{eqnarray} 
\beta(N) = \beta_{\infty} + a N^{-b}, 
\end{eqnarray}
where $\beta_\infty = 1.20(16)$, $a = -1.21(7)$, and $b = 0.10(4)$ if
the fit is restricted to systems of size $N>10^4$. Upon successively
excluding the smaller system sizes from the fit we find that the value
of $\beta_\infty$ approaches the expected mean-field value $\beta=1$
\cite{moore:00}.  For example, restricting the analysis to $N>2\times
10^4$ yields $\beta_\infty = 1.06(7)$, $a = -1.5(4)$, and $b = 0.16(4)$
[see the inset of Fig.~\ref{fig:SWCG_fss}(a)]. Note  that in the figure we
fixed $\beta_\infty=1$.

An additional estimate of the critical point can be obtained from the
position of the peaks of the finite-size susceptibility $\chi(p)$.  We
have located the individual peak positions $p_{\rm max}(N)$ by fitting a
polynomial of fifth order to the unscaled data curves. This is illustrated
in Fig.~\ref{fig:SWCG_fss}(b), where the main plot shows the raw data
with the respective fits and the inset shows the scaling behavior of the
peak positions, where a fit to the function
\begin{eqnarray}
p_{\rm max}(N) = p_{\rm max}^\infty + a N^{-b}
\end{eqnarray}
yields $p_{\rm max}^\infty = 0.2014(2)$, $a = 0.54(3)$, and $b =
0.341(7)$.  The value of $p_{\rm max}^\infty$ is in reasonable agreement
with the above estimate based on the analysis of the order parameter.
Furthermore, the numerical value of the critical point compares well with
an estimate $p_{\rm c}=0.201(1)$ obtained using a data-collapse analysis
(not shown).

Note that both estimates, $p_{\rm c}^\infty$ and $p_{\rm max}^\infty$,
are in reasonable agreement and are located significantly below the
threshold value $p_{\rm c}=0.38675(7)$ of the standard chimera graph.
Consequently, SWCGs provide a topology that is significantly less
vulnerable to random failures of qubits, i.e., while the standard
chimera graph exhibits a fragmentation threshold $f_{\rm c}=1-p_{\rm
c}\approx 0.62$ and thus allows for a random failure of approximately
$62\%$ of the qubits without losing large-scale connectivity, this value
increases to $f_{\rm c}\approx 0.80$ for the ensemble of SWCGs.
Finally, we note that the critical exponents for percolation on SWCGs
assume mean-field values when ${\mathcal O}(N)$ small-world bonds are
added, as demonstrated in the presented study.

Finally, note that chimera topologies are the archetypal architecture
used in current quantum annealers. While, from a point of view of
robustness, a fully connected topology would be desirable, a hardware
implementation seems not possible at present. To be precise, only a
finite number of fabrication layers for the chips are available. Having
a fully connected graph would require $O(N)$ layers, which is
prohibitive for current chip designs with $\sim 1000$ qubits. Given the
flux qubit structure used in current quantum annealing machines,
$K_{n,n}$-like topologies might be used for multiple upcoming
generations of these devices.

\section{Summary} 
\label{sec:summary}

We have performed numerical simulations to determine the bond- and
site-percolation thresholds on nonplanar, effectively two-dimensional
lattice graphs, where the elementary building blocks are complete
bipartite subgraphs $K_{n,n}$ ($n = 2, \ldots, 8$). The simulations
have been performed using a highly efficient percolation algorithm
\cite{ziff:00,ziff:01} based on a union-find data structure
\cite{cormen:01}. From a finite-size scaling analysis we have obtained
the critical points $p_{\rm c}$ and the three critical exponents $\nu$,
$\beta$, and $\gamma$, thus locating the critical bond- and
site-percolation thresholds and allowing us to verify that the
transition is in the two-dimensional percolation universality class. In
either case, the percolation threshold is a decreasing function of $n$
and our result suggests that in the asymptotic limit $p_{\rm c} \to 0$
as $n \to \infty$.

The particular choice of $n=4$ is the canonical chimera graph, i.e., the
hardware topology of the D-Wave quantum annealing device, developed at
D-Wave Systems Inc.~\cite{comment:d-wave}. The native (no embedding
required) benchmark (optimization) problem for the D-Wave device is an
Ising spin glass \cite{binder:86,hartmann:01} and recently, much effort
was put into the simulation of Ising spin glasses on the chimera
topology \cite{katzgraber:14,weigel:15,katzgraber:15}.  As discussed in
Ref.~\cite{zhu:15b}, the location of the site-percolation threshold is
crucial for the efficient and correct performance of cluster algorithms
designed to simulate spin-glass models on, e.g., the above graph
topology.

Finally, referring to the implementation of, e.g., the D-Wave chip and
adopting the point of view of network robustness and vulnerability, the
above results suggest that the native D-Wave design, as analyzed in
Secs.~\ref{subsec:res_bondPerc} and \ref{subsec:res_sitePerc}, allows
for a random failure of approximately $62\%$ of the qubits ($70\%$ of
the couplers) prior to losing large-scale connectivity on the chip.
Similarly, embedded problems that turn off a sizable fraction of
couplers randomly, might lead to loss of connectivity.  Bear in mind
that the above figures are valid in the asymptotic limit. In general,
for finite-sized graphs of no more than $10^3$ sites, finite-size
effects result in effective thresholds that differ slightly from the
asymptotic values quoted in Table \ref{tab:tab2}.  To illustrate this,
one might, e.g., define effective, system-size-dependent critical points
from the peak locations of the finite-size fluctuations $\chi$ (see
Sec.~\ref{par:orderParFluct}).  In this regard, for bond (site)
percolation on a lattice with $N=512$ sites we observe $p_{\chi-{\rm
max}}(N=512)\approx0.307$ [$p_{\chi-{\rm max}}(N=512)\approx0.408$],
i.e., shifting towards smaller values as $N\to \infty$.  Similarly, for
$N=1152$, $p_{\chi-{\rm max}}(N=1152)\approx0.304$ for bond percolation
and $p_{\chi-{\rm max}}(N=1152)\approx0.403$ for site percolation.
Finally, for the largest system sizes studied in this work,
$p_{\chi-{\rm max}}(N=131072)\approx0.296$ for bond percolation and
$p_{\chi-{\rm max}}(N=294912)\approx0.389$ for site percolation.
Although the asymptotic peaks seem to be located slightly above $p_{\rm
c}$ (within the superpercolating regime), this might nevertheless lead
to expect that the finite-size values of $p_{\rm c}$ for bond and site
percolation for the $N=1152$ chimera graph are within a $5\%$ interval
of the asymptotic critical point.

In addition, we have found that by extending the plain $K_{4,4}$-based
chimera graph using $N/2$ small-world bonds---thereby effectively
increasing the average vertex degree by one---the respective percolation
threshold decreases to $p_{\rm c}=0.207(4)$. Thus, small-world-extended
chimera graphs provide a topology that allows for a random failure of
approximately $80\%$ of the qubits before the large-scale connectivity of
the device is lost. As pointed out earlier, using topologies that have
higher connectivity, such as the above extended chimera graphs, might be
key in designing quantum annealing machines robust to random failures of
qubits and couplers. 

\begin{acknowledgments}

We would like to thank R.~Ziff for sharing his estimates of $p_{{\rm
c},2}$ with us, as well as fruitful discussions.  O.M.~thanks Zheng Zhu
for valuable discussions and comments as well as for critically reading
the manuscript.  H.G.K.~thanks Firas Hamze for comments and discussions
and acknowledges support from the National Science Foundation (Grant
No.~DMR-1151387). The research of H.G.K.~and O.M.~is based in part
upon work supported in part by the Office of the Director of National
Intelligence (ODNI), Intelligence Advanced Research Projects Activity
(IARPA), via MIT Lincoln Laboratory Air Force Contract
No.~FA8721-05-C-0002.  The views and conclusions contained herein are
those of the authors and should not be interpreted as necessarily
representing the official policies or endorsements, either expressed or
implied, of ODNI, IARPA, or the U.S.~Government.  The U.S.~Government is
authorized to reproduce and distribute reprints for Governmental purpose
notwithstanding any copyright annotation thereon.  M.A.N.~is supported
in part by a grant from Pacific Northwest National Laboratory (PNNL).

\end{acknowledgments}

\bibliography{refs.bib,comments.bib}

\begin{thebibliography}{56}
\expandafter\ifx\csname natexlab\endcsname\relax\def\natexlab#1{#1}\fi
\expandafter\ifx\csname bibnamefont\endcsname\relax
  \def\bibnamefont#1{#1}\fi
\expandafter\ifx\csname bibfnamefont\endcsname\relax
  \def\bibfnamefont#1{#1}\fi
\expandafter\ifx\csname citenamefont\endcsname\relax
  \def\citenamefont#1{#1}\fi
\expandafter\ifx\csname url\endcsname\relax
  \def\url#1{\texttt{#1}}\fi
\expandafter\ifx\csname urlprefix\endcsname\relax\def\urlprefix{URL }\fi
\providecommand{\bibinfo}[2]{#2}
\providecommand{\eprint}[2][]{\url{#2}}

\bibitem[{\citenamefont{{Broadbent} and {Hammersley}}(1957)}]{broadbent:57}
\bibinfo{author}{\bibfnamefont{S.~R.} \bibnamefont{{Broadbent}}}
  \bibnamefont{and} \bibinfo{author}{\bibfnamefont{J.~M.}
  \bibnamefont{{Hammersley}}}, \emph{\bibinfo{title}{{{Percolation processes.
  I. Crystals and Mazes}}}}, \bibinfo{journal}{Proceedings of the Cambridge
  Philosophical Society} \textbf{\bibinfo{volume}{53}}, \bibinfo{pages}{629}
  (\bibinfo{year}{1957}).

\bibitem[{\citenamefont{{Hammersley, J. M.}}({1957})}]{hammersley:57}
\bibinfo{author}{\bibnamefont{{Hammersley, J. M.}}},
  \emph{\bibinfo{title}{{Percolation Processes: Lower Bounds for the Critical
  Probability}}}, \bibinfo{journal}{{Ann. Math. Statist.}}
  \textbf{\bibinfo{volume}{{28}}}, \bibinfo{pages}{{790}}
  (\bibinfo{year}{{1957}}).

\bibitem[{\citenamefont{{Hammersley}}(1957)}]{hammersley:57a}
\bibinfo{author}{\bibfnamefont{J.~M.} \bibnamefont{{Hammersley}}},
  \emph{\bibinfo{title}{{{Percolation processes. II. The connective
  constant}}}}, \bibinfo{journal}{Proceedings of the Cambridge Philosophical
  Society} \textbf{\bibinfo{volume}{53}}, \bibinfo{pages}{642}
  (\bibinfo{year}{1957}).

\bibitem[{\citenamefont{Fisher}(1961)}]{fisher:61a}
\bibinfo{author}{\bibfnamefont{M.~E.} \bibnamefont{Fisher}},
  \emph{\bibinfo{title}{{Critical Probabilities for Cluster Size and
  Percolation Problems}}}, \bibinfo{journal}{J. Math. Phys.}
  \textbf{\bibinfo{volume}{2}}, \bibinfo{pages}{620} (\bibinfo{year}{1961}).

\bibitem[{\citenamefont{Bunde et~al.}(1991)\citenamefont{Bunde, Maass, and
  Ingram}}]{bunde:91}
\bibinfo{author}{\bibfnamefont{A.}~\bibnamefont{Bunde}},
  \bibinfo{author}{\bibfnamefont{P.}~\bibnamefont{Maass}}, \bibnamefont{and}
  \bibinfo{author}{\bibfnamefont{M.~D.} \bibnamefont{Ingram}},
  \emph{\bibinfo{title}{{Diffusion Limited Percolation: A Model for Transport
  in Ionic Glasses}}}, \bibinfo{journal}{Berichte der Bunsengesellschaft f\"ur
  physikalische Chemie} \textbf{\bibinfo{volume}{95}}, \bibinfo{pages}{977}
  (\bibinfo{year}{1991}).

\bibitem[{\citenamefont{Pfeiffer and Rieger}(2002)}]{pfeiffer:02}
\bibinfo{author}{\bibfnamefont{F.~O.} \bibnamefont{Pfeiffer}} \bibnamefont{and}
  \bibinfo{author}{\bibfnamefont{H.}~\bibnamefont{Rieger}},
  \emph{\bibinfo{title}{{{Superconductor-to-normal phase transition in a vortex
  glass model: numerical evidence for a new percolation universality class}}}},
  \bibinfo{journal}{J. Phys. Cond. Mat.} \textbf{\bibinfo{volume}{14}},
  \bibinfo{pages}{2361} (\bibinfo{year}{2002}).

\bibitem[{\citenamefont{Pfeiffer and Rieger}(2003)}]{pfeiffer:03}
\bibinfo{author}{\bibfnamefont{F.~O.} \bibnamefont{Pfeiffer}} \bibnamefont{and}
  \bibinfo{author}{\bibfnamefont{H.}~\bibnamefont{Rieger}},
  \emph{\bibinfo{title}{{{Critical properties of loop percolation models with
  optimization constraints}}}}, \bibinfo{journal}{Phys. Rev. E}
  \textbf{\bibinfo{volume}{67}}, \bibinfo{pages}{056113}
  (\bibinfo{year}{2003}).

\bibitem[{\citenamefont{Stauffer}(1979)}]{stauffer:79}
\bibinfo{author}{\bibfnamefont{D.}~\bibnamefont{Stauffer}},
  \emph{\bibinfo{title}{{Scaling theory of percolation clusters}}},
  \bibinfo{journal}{Phys. Rep.} \textbf{\bibinfo{volume}{54}},
  \bibinfo{pages}{1} (\bibinfo{year}{1979}).

\bibitem[{\citenamefont{Stauffer and Aharony}(1994)}]{stauffer:94}
\bibinfo{author}{\bibfnamefont{D.}~\bibnamefont{Stauffer}} \bibnamefont{and}
  \bibinfo{author}{\bibfnamefont{A.}~\bibnamefont{Aharony}},
  \emph{\bibinfo{title}{{{Introduction to Percolation Theory}}}}
  (\bibinfo{publisher}{Taylor and Francis, London}, \bibinfo{year}{1994}).

\bibitem[{\citenamefont{Essam and Fisher}(1970)}]{essam:70}
\bibinfo{author}{\bibfnamefont{J.~W.} \bibnamefont{Essam}} \bibnamefont{and}
  \bibinfo{author}{\bibfnamefont{M.~E.} \bibnamefont{Fisher}},
  \emph{\bibinfo{title}{{{Some Basic Definitions in Graph Theory}}}},
  \bibinfo{journal}{Rev. Mod. Phys.} \textbf{\bibinfo{volume}{42}},
  \bibinfo{pages}{271} (\bibinfo{year}{1970}).

\bibitem[{\citenamefont{Yeomans}(1992)}]{yeomans:92}
\bibinfo{author}{\bibfnamefont{J.~M.} \bibnamefont{Yeomans}},
  \emph{\bibinfo{title}{{Statistical Mechanics of Phase Transitions}}}
  (\bibinfo{publisher}{Oxford University Press}, \bibinfo{address}{Oxford},
  \bibinfo{year}{1992}).

\bibitem[{\citenamefont{Becker and Ziff}(2009)}]{becker:09}
\bibinfo{author}{\bibfnamefont{A.~M.} \bibnamefont{Becker}} \bibnamefont{and}
  \bibinfo{author}{\bibfnamefont{R.~M.} \bibnamefont{Ziff}},
  \emph{\bibinfo{title}{{{Percolation thresholds on two-dimensional Voronoi
  networks and Delaunay triangulations}}}}, \bibinfo{journal}{Phys. Rev. E}
  \textbf{\bibinfo{volume}{80}}, \bibinfo{pages}{041101}
  (\bibinfo{year}{2009}).

\bibitem[{\citenamefont{{Saberi}}(2015)}]{saberi:15}
\bibinfo{author}{\bibfnamefont{A.~A.} \bibnamefont{{Saberi}}},
  \emph{\bibinfo{title}{{Recent advances in percolation theory and its
  applications}}}, \bibinfo{journal}{Physics Reports}
  \textbf{\bibinfo{volume}{578}}, \bibinfo{pages}{1} (\bibinfo{year}{2015}).

\bibitem[{\citenamefont{Newman et~al.}(2001)\citenamefont{Newman, Strogatz, and
  Watts}}]{newman:01a}
\bibinfo{author}{\bibfnamefont{M.~E.~J.} \bibnamefont{Newman}},
  \bibinfo{author}{\bibfnamefont{S.~H.} \bibnamefont{Strogatz}},
  \bibnamefont{and} \bibinfo{author}{\bibfnamefont{D.~J.} \bibnamefont{Watts}},
  \emph{\bibinfo{title}{{Random graphs with arbitrary degree distributions and
  their applications}}}, \bibinfo{journal}{Phys. Rev. E}
  \textbf{\bibinfo{volume}{64}}, \bibinfo{pages}{026118}
  (\bibinfo{year}{2001}).

\bibitem[{\citenamefont{Callaway et~al.}(2000)\citenamefont{Callaway, Newman,
  Strogatz, and Watts}}]{callaway:00}
\bibinfo{author}{\bibfnamefont{D.~S.} \bibnamefont{Callaway}},
  \bibinfo{author}{\bibfnamefont{M.~E.~J.} \bibnamefont{Newman}},
  \bibinfo{author}{\bibfnamefont{S.~H.} \bibnamefont{Strogatz}},
  \bibnamefont{and} \bibinfo{author}{\bibfnamefont{D.~J.} \bibnamefont{Watts}},
  \emph{\bibinfo{title}{{Network Robustness and Fragility: Percolation on
  Random Graphs}}}, \bibinfo{journal}{Phys. Rev. Lett.}
  \textbf{\bibinfo{volume}{85}}, \bibinfo{pages}{5468} (\bibinfo{year}{2000}).

\bibitem[{\citenamefont{Cormen et~al.}(2001)\citenamefont{Cormen, Leiserson,
  Rivest, and Stein}}]{cormen:01}
\bibinfo{author}{\bibfnamefont{T.~H.} \bibnamefont{Cormen}},
  \bibinfo{author}{\bibfnamefont{C.~E.} \bibnamefont{Leiserson}},
  \bibinfo{author}{\bibfnamefont{R.~L.} \bibnamefont{Rivest}},
  \bibnamefont{and} \bibinfo{author}{\bibfnamefont{C.}~\bibnamefont{Stein}},
  \emph{\bibinfo{title}{{Introduction to Algorithms, 2nd edition}}}
  (\bibinfo{publisher}{MIT Press}, \bibinfo{address}{Cambridge, MA},
  \bibinfo{year}{2001}).

\bibitem[{\citenamefont{Newman and Ziff}(2000)}]{ziff:00}
\bibinfo{author}{\bibfnamefont{M.~E.~J.} \bibnamefont{Newman}}
  \bibnamefont{and} \bibinfo{author}{\bibfnamefont{R.}~\bibnamefont{Ziff}},
  \emph{\bibinfo{title}{{{Efficient Monte Carlo Algorithm and High-Precision
  Results for Percolation}}}}, \bibinfo{journal}{Phys. Rev. Lett.}
  \textbf{\bibinfo{volume}{85}}, \bibinfo{pages}{4104} (\bibinfo{year}{2000}).

\bibitem[{\citenamefont{Newman and Ziff}(2001)}]{ziff:01}
\bibinfo{author}{\bibfnamefont{M.~E.~J.} \bibnamefont{Newman}}
  \bibnamefont{and} \bibinfo{author}{\bibfnamefont{R.}~\bibnamefont{Ziff}},
  \emph{\bibinfo{title}{{{Fast Monte Carlo algorithm for site or bond
  percolation}}}}, \bibinfo{journal}{Phys. Rev. E}
  \textbf{\bibinfo{volume}{64}}, \bibinfo{pages}{016706}
  (\bibinfo{year}{2001}).

\bibitem[{\citenamefont{Mertens and Moore}(2012)}]{mertens:12}
\bibinfo{author}{\bibfnamefont{S.}~\bibnamefont{Mertens}} \bibnamefont{and}
  \bibinfo{author}{\bibfnamefont{C.}~\bibnamefont{Moore}},
  \emph{\bibinfo{title}{Continuum percolation thresholds in two dimensions}},
  \bibinfo{journal}{Phys. Rev. E} \textbf{\bibinfo{volume}{86}},
  \bibinfo{pages}{061109} (\bibinfo{year}{2012}).

\bibitem[{\citenamefont{Bunyk et~al.}(2014)\citenamefont{Bunyk, Hoskinson,
  Johnson, Tolkacheva, Altomare, Berkley, Harris, Hilton, Lanting, and
  Whittaker}}]{bunyk:14}
\bibinfo{author}{\bibfnamefont{P.}~\bibnamefont{Bunyk}},
  \bibinfo{author}{\bibfnamefont{E.}~\bibnamefont{Hoskinson}},
  \bibinfo{author}{\bibfnamefont{M.~W.} \bibnamefont{Johnson}},
  \bibinfo{author}{\bibfnamefont{E.}~\bibnamefont{Tolkacheva}},
  \bibinfo{author}{\bibfnamefont{F.}~\bibnamefont{Altomare}},
  \bibinfo{author}{\bibfnamefont{A.~J.} \bibnamefont{Berkley}},
  \bibinfo{author}{\bibfnamefont{R.}~\bibnamefont{Harris}},
  \bibinfo{author}{\bibfnamefont{J.~P.} \bibnamefont{Hilton}},
  \bibinfo{author}{\bibfnamefont{T.}~\bibnamefont{Lanting}}, \bibnamefont{and}
  \bibinfo{author}{\bibfnamefont{J.}~\bibnamefont{Whittaker}},
  \emph{\bibinfo{title}{{Architectural Considerations in the Design of a
  Superconducting Quantum Annealing Processor}}}, \bibinfo{journal}{IEEE Trans.
  Appl. Supercond.} \textbf{\bibinfo{volume}{24}}, \bibinfo{pages}{1}
  (\bibinfo{year}{2014}).

\bibitem[{com({\natexlab{a}})}]{comment:d-wave}
\urlprefix\url{http://www.dwavesys.com}.

\bibitem[{\citenamefont{Bian et~al.}(2014)\citenamefont{Bian, Chudak, Israel,
  Lackey, Macready, and Roy}}]{bian:14}
\bibinfo{author}{\bibfnamefont{Z.}~\bibnamefont{Bian}},
  \bibinfo{author}{\bibfnamefont{F.}~\bibnamefont{Chudak}},
  \bibinfo{author}{\bibfnamefont{R.}~\bibnamefont{Israel}},
  \bibinfo{author}{\bibfnamefont{B.}~\bibnamefont{Lackey}},
  \bibinfo{author}{\bibfnamefont{W.~G.} \bibnamefont{Macready}},
  \bibnamefont{and} \bibinfo{author}{\bibfnamefont{A.}~\bibnamefont{Roy}},
  \emph{\bibinfo{title}{{Discrete optimization using Quantum Annealing on
  sparse Ising models}}}, \bibinfo{journal}{Frontiers in Physics}
  \textbf{\bibinfo{volume}{2}} (\bibinfo{year}{2014}).

\bibitem[{\citenamefont{Klymko et~al.}(2014)\citenamefont{Klymko, Sullivan, and
  Humble}}]{klymko:14}
\bibinfo{author}{\bibfnamefont{C.}~\bibnamefont{Klymko}},
  \bibinfo{author}{\bibfnamefont{B.~D.} \bibnamefont{Sullivan}},
  \bibnamefont{and} \bibinfo{author}{\bibfnamefont{T.~S.}
  \bibnamefont{Humble}}, \emph{\bibinfo{title}{Adiabatic quantum programming:
  minor embedding with hard faults}}, \bibinfo{journal}{Quant. Inf. Proc.}
  \textbf{\bibinfo{volume}{13}}, \bibinfo{pages}{709} (\bibinfo{year}{2014}).

\bibitem[{\citenamefont{Albert et~al.}(2000)\citenamefont{Albert, Jeong, and
  Barab{\'a}si}}]{albert:00}
\bibinfo{author}{\bibfnamefont{R.}~\bibnamefont{Albert}},
  \bibinfo{author}{\bibfnamefont{H.}~\bibnamefont{Jeong}}, \bibnamefont{and}
  \bibinfo{author}{\bibfnamefont{A.}~\bibnamefont{Barab{\'a}si}},
  \emph{\bibinfo{title}{{Error and attack tolerance of complex networks}}},
  \bibinfo{journal}{Nature} \textbf{\bibinfo{volume}{406}},
  \bibinfo{pages}{378} (\bibinfo{year}{2000}).

\bibitem[{com({\natexlab{b}})}]{comment:vinci}
\bibinfo{note}{Note that in Ref.~\cite{vinci:15} the percolation properties of
  a two-level-grid minor embedded into chimera were studied within the context
  of quantum annealing corrections.}

\bibitem[{com({\natexlab{c}})}]{comment:native}
\bibinfo{note}{Within this context, {\em native} refers to a problem that uses
  all physical qubits on the chip as logical qubits. An embedded problem, for
  example, might require multiple physical qubits to encode one logical qubit
  or interaction between two qubits that are not nearest neighbors on the
  lattice.}

\bibitem[{\citenamefont{Binder and Young}(1986)}]{binder:86}
\bibinfo{author}{\bibfnamefont{K.}~\bibnamefont{Binder}} \bibnamefont{and}
  \bibinfo{author}{\bibfnamefont{A.~P.} \bibnamefont{Young}},
  \emph{\bibinfo{title}{{Spin Glasses: Experimental Facts, Theoretical Concepts
  and Open Questions}}}, \bibinfo{journal}{Rev. Mod. Phys.}
  \textbf{\bibinfo{volume}{58}}, \bibinfo{pages}{801} (\bibinfo{year}{1986}).

\bibitem[{\citenamefont{Stein and Newman}(2013)}]{stein:13}
\bibinfo{author}{\bibfnamefont{D.~L.} \bibnamefont{Stein}} \bibnamefont{and}
  \bibinfo{author}{\bibfnamefont{C.~M.} \bibnamefont{Newman}},
  \emph{\bibinfo{title}{{Spin Glasses and Complexity}}}, Primers in Complex
  Systems (\bibinfo{publisher}{Princeton University Press},
  \bibinfo{year}{2013}).

\bibitem[{\citenamefont{Katzgraber et~al.}(2014)\citenamefont{Katzgraber,
  Hamze, and Andrist}}]{katzgraber:14}
\bibinfo{author}{\bibfnamefont{H.~G.} \bibnamefont{Katzgraber}},
  \bibinfo{author}{\bibfnamefont{F.}~\bibnamefont{Hamze}}, \bibnamefont{and}
  \bibinfo{author}{\bibfnamefont{R.~S.} \bibnamefont{Andrist}},
  \emph{\bibinfo{title}{{Glassy Chimeras Could Be Blind to Quantum Speedup:
  Designing Better Benchmarks for Quantum Annealing Machines}}},
  \bibinfo{journal}{Phys. Rev. X} \textbf{\bibinfo{volume}{4}},
  \bibinfo{pages}{021008} (\bibinfo{year}{2014}).

\bibitem[{\citenamefont{Katzgraber et~al.}(2015)\citenamefont{Katzgraber,
  Hamze, Zhu, Ochoa, and Munoz-Bauza}}]{katzgraber:15}
\bibinfo{author}{\bibfnamefont{H.~G.} \bibnamefont{Katzgraber}},
  \bibinfo{author}{\bibfnamefont{F.}~\bibnamefont{Hamze}},
  \bibinfo{author}{\bibfnamefont{Z.}~\bibnamefont{Zhu}},
  \bibinfo{author}{\bibfnamefont{A.~J.} \bibnamefont{Ochoa}}, \bibnamefont{and}
  \bibinfo{author}{\bibfnamefont{H.}~\bibnamefont{Munoz-Bauza}},
  \emph{\bibinfo{title}{{Seeking Quantum Speedup Through Spin Glasses: The
  Good, the Bad, and the Ugly}}}, \bibinfo{journal}{Phys. Rev. X}
  \textbf{\bibinfo{volume}{5}}, \bibinfo{pages}{031026} (\bibinfo{year}{2015}).

\bibitem[{\citenamefont{{Zhu} et~al.}(2015)\citenamefont{{Zhu}, {Ochoa}, and
  {Katzgraber}}}]{zhu:15b}
\bibinfo{author}{\bibfnamefont{Z.}~\bibnamefont{{Zhu}}},
  \bibinfo{author}{\bibfnamefont{A.~J.} \bibnamefont{{Ochoa}}},
  \bibnamefont{and} \bibinfo{author}{\bibfnamefont{H.~G.}
  \bibnamefont{{Katzgraber}}}, \emph{\bibinfo{title}{{{Efficient Cluster
  Algorithm for Spin Glasses in Any Space Dimension}}}},
  \bibinfo{journal}{Phys. Rev. Lett.} \textbf{\bibinfo{volume}{115}},
  \bibinfo{pages}{077201} (\bibinfo{year}{2015}).

\bibitem[{\citenamefont{Hen et~al.}(2015)\citenamefont{Hen, Job, Albash,
  R{\o}nnow, Troyer, and Lidar}}]{hen:15a}
\bibinfo{author}{\bibfnamefont{I.}~\bibnamefont{Hen}},
  \bibinfo{author}{\bibfnamefont{J.}~\bibnamefont{Job}},
  \bibinfo{author}{\bibfnamefont{T.}~\bibnamefont{Albash}},
  \bibinfo{author}{\bibfnamefont{T.~F.} \bibnamefont{R{\o}nnow}},
  \bibinfo{author}{\bibfnamefont{M.}~\bibnamefont{Troyer}}, \bibnamefont{and}
  \bibinfo{author}{\bibfnamefont{D.~A.} \bibnamefont{Lidar}},
  \emph{\bibinfo{title}{{Probing for quantum speedup in spin-glass problems
  with planted solutions}}}, \bibinfo{journal}{Phys. Rev. A}
  \textbf{\bibinfo{volume}{92}}, \bibinfo{pages}{042325}
  (\bibinfo{year}{2015}).

\bibitem[{\citenamefont{King}(2015)}]{king:15}
\bibinfo{author}{\bibfnamefont{A.~D.} \bibnamefont{King}},
  \emph{\bibinfo{title}{{{Performance of a quantum annealer on range-limited
  constraint satisfaction problems}}}} (\bibinfo{year}{2015}),
  \bibinfo{note}{arXiv:1502.02098}.

\bibitem[{hal(2015)}]{hall:15}
\emph{\bibinfo{title}{{{A Study of Spanning Trees on a D-Wave Quantum
  Computer}}}}, \bibinfo{journal}{{Physics Procedia}}
  \textbf{\bibinfo{volume}{68}}, \bibinfo{pages}{56} (\bibinfo{year}{2015}).

\bibitem[{\citenamefont{{Boixo} et~al.}(2014)\citenamefont{{Boixo},
  {R{\o}nnow}, {Isakov}, {Wang}, {Wecker}, {Lidar}, {Martinis}, and
  {Troyer}}}]{boixo:14}
\bibinfo{author}{\bibfnamefont{S.}~\bibnamefont{{Boixo}}},
  \bibinfo{author}{\bibfnamefont{T.~F.} \bibnamefont{{R{\o}nnow}}},
  \bibinfo{author}{\bibfnamefont{S.~V.} \bibnamefont{{Isakov}}},
  \bibinfo{author}{\bibfnamefont{Z.}~\bibnamefont{{Wang}}},
  \bibinfo{author}{\bibfnamefont{D.}~\bibnamefont{{Wecker}}},
  \bibinfo{author}{\bibfnamefont{D.~A.} \bibnamefont{{Lidar}}},
  \bibinfo{author}{\bibfnamefont{J.~M.} \bibnamefont{{Martinis}}},
  \bibnamefont{and} \bibinfo{author}{\bibfnamefont{M.}~\bibnamefont{{Troyer}}},
  \emph{\bibinfo{title}{{Evidence for quantum annealing with more than one
  hundred qubits}}}, \bibinfo{journal}{Nat. Phys,}
  \textbf{\bibinfo{volume}{10}}, \bibinfo{pages}{218} (\bibinfo{year}{2014}).

\bibitem[{\citenamefont{Binder}(1981{\natexlab{a}})}]{binder:81}
\bibinfo{author}{\bibfnamefont{K.}~\bibnamefont{Binder}},
  \emph{\bibinfo{title}{Critical properties from {M}onte {C}arlo coarse
  graining and renormalization}}, \bibinfo{journal}{Phys. Rev. Lett.}
  \textbf{\bibinfo{volume}{47}}, \bibinfo{pages}{693}
  (\bibinfo{year}{1981}{\natexlab{a}}).

\bibitem[{com({\natexlab{d}})}]{comment:essam}
\bibinfo{note}{Note that in Ref.~\cite{essam:70} these bipartite graphs are
  referred to as ``bichromatic.''}.

\bibitem[{com({\natexlab{e}})}]{comment:dream}
\bibinfo{note}{Needless to mention, the expectation that the current chip
  topology will scale to hundreds of thousands of qubits requires an
  unhealthily large amount of wishful thinking. This estimate does not take
  into account, for example, fabrication limitations, or the effects of qubit
  noise that are amplified the more qubits are added to the system
  \cite{zhu:16,perdomo:15,perdomo:15a}.}

\bibitem[{\citenamefont{Binder and Heermann}(2002)}]{binder:02a}
\bibinfo{author}{\bibfnamefont{K.}~\bibnamefont{Binder}} \bibnamefont{and}
  \bibinfo{author}{\bibfnamefont{D.~W.} \bibnamefont{Heermann}},
  \emph{\bibinfo{title}{{Monte Carlo simulation in statistical physics: an
  introduction (4th ed.)}}} (\bibinfo{publisher}{Springer (Springer series in
  solid-state science)}, \bibinfo{year}{2002}).

\bibitem[{\citenamefont{Melchert}(2009)}]{melchert-autoscale:09}
\bibinfo{author}{\bibfnamefont{O.}~\bibnamefont{Melchert}},
  \emph{\bibinfo{title}{{autoScale.py - A program for automatic finite-size
  scaling analyses: A user's guide}}}, \bibinfo{journal}{Preprint:
  arXiv:0910.5403v1}  (\bibinfo{year}{2009}), \bibinfo{note}{the source-code of
  autoScale.py and the raw-data for an illustrative example can be downloaded
  toghether with the source-files of the preprint at
  http://arxiv.org/abs/0910.5403 by choosing the download-option {\tt Other
  formats}.}

\bibitem[{\citenamefont{Sorge}(2015)}]{comment:sorge}
\bibinfo{author}{\bibfnamefont{A.}~\bibnamefont{Sorge}},
  \emph{\bibinfo{title}{pyfssa: v0.2.0}} (\bibinfo{year}{2015}),
  \bibinfo{note}{{\texttt{pyfssa} is a scientific Python package for
  algorithmic finite-size scaling analysis at phase transitions.}}

\bibitem[{\citenamefont{Houdayer and Hartmann}(2004)}]{houdayer:04}
\bibinfo{author}{\bibfnamefont{J.}~\bibnamefont{Houdayer}} \bibnamefont{and}
  \bibinfo{author}{\bibfnamefont{A.~K.} \bibnamefont{Hartmann}},
  \emph{\bibinfo{title}{Low temperature behavior of two-dimensional {G}aussian
  {I}sing spin glasses}}, \bibinfo{journal}{Phys. Rev. B}
  \textbf{\bibinfo{volume}{70}}, \bibinfo{pages}{014418}
  (\bibinfo{year}{2004}).

\bibitem[{\citenamefont{Binder}(1981{\natexlab{b}})}]{binder:81b}
\bibinfo{author}{\bibfnamefont{K.}~\bibnamefont{Binder}},
  \emph{\bibinfo{title}{Finite size scaling analysis of {I}sing model block
  distribution functions}}, \bibinfo{journal}{Z. Phys. B}
  \textbf{\bibinfo{volume}{43}}, \bibinfo{pages}{119}
  (\bibinfo{year}{1981}{\natexlab{b}}).

\bibitem[{com({\natexlab{f}})}]{comment:S}
\bibinfo{note}{The numerical value of $S$ measures the mean-square distance of
  the data points to the master scaling curve described by the scaling
  function, in units of the standard error \cite{houdayer:04}}.

\bibitem[{\citenamefont{Wierman}(2002)}]{wierman:02}
\bibinfo{author}{\bibfnamefont{J.~C.} \bibnamefont{Wierman}},
  \emph{\bibinfo{title}{{Percolation threshold is not a decreasing function of
  the average coordination number}}}, \bibinfo{journal}{Phys. Rev. E}
  \textbf{\bibinfo{volume}{66}}, \bibinfo{pages}{046125}
  (\bibinfo{year}{2002}).

\bibitem[{com({\natexlab{g}})}]{comment:rziff_pc2}
\bibinfo{note}{R.~Ziff (private communication) drew our attention to his
  independent estimates of $p_{{\rm c},2} \approx 0.42776$ (bond percolation)
  and $p_{{\rm c},2} \approx 0.51298$ (site percolation) on the $K_{2,2}$
  chimera lattice, consistent with the values quoted in Table \ref{tab:tab2}.}

\bibitem[{\citenamefont{Guclu et~al.}(2006)\citenamefont{Guclu, Korniss,
  Novotny, Toroczkai, and R{\'a}cz}}]{guclu:06}
\bibinfo{author}{\bibfnamefont{H.}~\bibnamefont{Guclu}},
  \bibinfo{author}{\bibfnamefont{G.}~\bibnamefont{Korniss}},
  \bibinfo{author}{\bibfnamefont{M.~A.} \bibnamefont{Novotny}},
  \bibinfo{author}{\bibfnamefont{Z.}~\bibnamefont{Toroczkai}},
  \bibnamefont{and} \bibinfo{author}{\bibfnamefont{Z.}~\bibnamefont{R{\'a}cz}},
  \emph{\bibinfo{title}{{Synchronization landscapes in small-world-connected
  computer networks}}}, \bibinfo{journal}{Phys. Rev. E}
  \textbf{\bibinfo{volume}{73}}, \bibinfo{pages}{066115}
  (\bibinfo{year}{2006}).

\bibitem[{com({\natexlab{h}})}]{comment:odd}
\bibinfo{note}{If $N$ is odd, one qubit does not have a small-world bond.}

\bibitem[{\citenamefont{Nachmias}(2009)}]{nachmias:09}
\bibinfo{author}{\bibfnamefont{A.}~\bibnamefont{Nachmias}},
  \emph{\bibinfo{title}{{Mean-field conditions for percolation on finite
  graphs}}}, \bibinfo{journal}{Geometric and Functional Analysis}
  \textbf{\bibinfo{volume}{19}}, \bibinfo{pages}{1171} (\bibinfo{year}{2009}).

\bibitem[{\citenamefont{Moore and Newman}(2000)}]{moore:00}
\bibinfo{author}{\bibfnamefont{C.}~\bibnamefont{Moore}} \bibnamefont{and}
  \bibinfo{author}{\bibfnamefont{M.~E.~J.} \bibnamefont{Newman}},
  \emph{\bibinfo{title}{{Exact solution of site and bond percolation on
  small-world networks}}}, \bibinfo{journal}{Phys. Rev. E}
  \textbf{\bibinfo{volume}{62}}, \bibinfo{pages}{7059} (\bibinfo{year}{2000}).

\bibitem[{\citenamefont{Hartmann and Rieger}(2001)}]{hartmann:01}
\bibinfo{author}{\bibfnamefont{A.~K.} \bibnamefont{Hartmann}} \bibnamefont{and}
  \bibinfo{author}{\bibfnamefont{H.}~\bibnamefont{Rieger}},
  \emph{\bibinfo{title}{Optimization Algorithms in Physics}}
  (\bibinfo{publisher}{Wiley-VCH}, \bibinfo{address}{Berlin},
  \bibinfo{year}{2001}).

\bibitem[{\citenamefont{Weigel et~al.}(2015)\citenamefont{Weigel, Katzgraber,
  Machta, Hamze, Andrist, and {Octomore Collaboration}}}]{weigel:15}
\bibinfo{author}{\bibfnamefont{M.}~\bibnamefont{Weigel}},
  \bibinfo{author}{\bibfnamefont{H.~G.} \bibnamefont{Katzgraber}},
  \bibinfo{author}{\bibfnamefont{J.}~\bibnamefont{Machta}},
  \bibinfo{author}{\bibfnamefont{F.}~\bibnamefont{Hamze}},
  \bibinfo{author}{\bibfnamefont{R.~S.} \bibnamefont{Andrist}},
  \bibnamefont{and} \bibinfo{author}{\bibnamefont{{Octomore Collaboration}}},
  \emph{\bibinfo{title}{{Erratum: Glassy Chimeras could be blind to quantum
  speedup: Designing better benchmarks for quantum annealing machines [Phys.
  Rev. X 4, 021008 (2014)]}}}, \bibinfo{journal}{Phys. Rev. X}
  \textbf{\bibinfo{volume}{5}}, \bibinfo{pages}{019901} (\bibinfo{year}{2015}).

\bibitem[{\citenamefont{Vinci et~al.}(2015)\citenamefont{Vinci, Albash,
  Paz-Silva, Hen, and Lidar}}]{vinci:15}
\bibinfo{author}{\bibfnamefont{W.}~\bibnamefont{Vinci}},
  \bibinfo{author}{\bibfnamefont{T.}~\bibnamefont{Albash}},
  \bibinfo{author}{\bibfnamefont{G.}~\bibnamefont{Paz-Silva}},
  \bibinfo{author}{\bibfnamefont{I.}~\bibnamefont{Hen}}, \bibnamefont{and}
  \bibinfo{author}{\bibfnamefont{D.~A.} \bibnamefont{Lidar}},
  \emph{\bibinfo{title}{Quantum annealing correction with minor embedding}},
  \bibinfo{journal}{Phys. Rev. A} \textbf{\bibinfo{volume}{92}},
  \bibinfo{pages}{042310} (\bibinfo{year}{2015}).

\bibitem[{\citenamefont{Zhu et~al.}(2016)\citenamefont{Zhu, Ochoa, Hamze,
  Schnabel, and Katzgraber}}]{zhu:16}
\bibinfo{author}{\bibfnamefont{Z.}~\bibnamefont{Zhu}},
  \bibinfo{author}{\bibfnamefont{A.~J.} \bibnamefont{Ochoa}},
  \bibinfo{author}{\bibfnamefont{F.}~\bibnamefont{Hamze}},
  \bibinfo{author}{\bibfnamefont{S.}~\bibnamefont{Schnabel}}, \bibnamefont{and}
  \bibinfo{author}{\bibfnamefont{H.~G.} \bibnamefont{Katzgraber}},
  \emph{\bibinfo{title}{{{Best-case performance of quantum annealers on native
  spin-glass benchmarks: How chaos can affect success probabilities}}}},
  \bibinfo{journal}{Phys. Rev. A} \textbf{\bibinfo{volume}{93}},
  \bibinfo{pages}{012317} (\bibinfo{year}{2016}).

\bibitem[{\citenamefont{{Perdomo-Ortiz}
  et~al.}(2015{\natexlab{a}})\citenamefont{{Perdomo-Ortiz}, {O'Gorman},
  {Fluegemann}, {Biswas}, and {Smelyanskiy}}}]{perdomo:15}
\bibinfo{author}{\bibfnamefont{A.}~\bibnamefont{{Perdomo-Ortiz}}},
  \bibinfo{author}{\bibfnamefont{B.}~\bibnamefont{{O'Gorman}}},
  \bibinfo{author}{\bibfnamefont{J.}~\bibnamefont{{Fluegemann}}},
  \bibinfo{author}{\bibfnamefont{R.}~\bibnamefont{{Biswas}}}, \bibnamefont{and}
  \bibinfo{author}{\bibfnamefont{V.~N.} \bibnamefont{{Smelyanskiy}}},
  \emph{\bibinfo{title}{{Determination and correction of persistent biases in
  quantum annealers}}} (\bibinfo{year}{2015}{\natexlab{a}}),
  \bibinfo{note}{(arXiv:quant-phys/1503.05679)}.

\bibitem[{\citenamefont{{Perdomo-Ortiz}
  et~al.}(2015{\natexlab{b}})\citenamefont{{Perdomo-Ortiz}, {Fluegemann},
  {Biswas}, and {Smelyanskiy}}}]{perdomo:15a}
\bibinfo{author}{\bibfnamefont{A.}~\bibnamefont{{Perdomo-Ortiz}}},
  \bibinfo{author}{\bibfnamefont{J.}~\bibnamefont{{Fluegemann}}},
  \bibinfo{author}{\bibfnamefont{R.}~\bibnamefont{{Biswas}}}, \bibnamefont{and}
  \bibinfo{author}{\bibfnamefont{V.~N.} \bibnamefont{{Smelyanskiy}}},
  \emph{\bibinfo{title}{{{A Performance Estimator for Quantum Annealers: Gauge
  selection and Parameter Setting}}}} (\bibinfo{year}{2015}{\natexlab{b}}),
  \bibinfo{note}{(arXiv:quant-phys/1503.01083)}.

\end{thebibliography}

\end{document}